\documentclass[twocolumn]{aastex63}

\usepackage{graphicx}
\usepackage{amsmath,amsthm,amssymb}
\usepackage{mathtools}
\newcommand{\eqb}{\begin{eqnarray}}
\newcommand{\eqe}{\end{eqnarray}}
\newcommand{\diff}{{\rm d}}
\newcommand{\betaw}{\beta_{\rm w}}
\newcommand{\gammaw}{\gamma_{\rm w}}
\newcommand{\ionfraction}{\eta_{\rm ion}}

\newcommand{\ppars}{p_{\| s}}
\newcommand{\ppare}{p_{\|{\rm e}}}
\newcommand{\pperp}{p_\bot}
\newcommand{\pperps}{\left|p_{\bot s}\right|}
\newcommand{\pperpeq}{p_{\bot{\rm eq}}}
\newcommand{\kappaep}{\kappa_{\rm ep}}

\accepted{for publication in the Astrophysical Journal}
\begin{document}

\title{Inductive acceleration of ions in Poynting-flux dominated outflows}

\author[0000-0002-9859-0496]{John~G.~Kirk}
\affiliation{Max-Planck-Institut f\"ur Kernphysik, Postfach 103980, 69029 Heidelberg, Germany}

\author{Gwenael~Giacinti}
\affiliation{Max-Planck-Institut f\"ur Kernphysik, Postfach 103980, 69029 Heidelberg, Germany}

\shortauthors{Kirk \& Giacinti}
\shorttitle{Inductive particle acceleration}

\begin{abstract}
  Two-fluid (electron-positron) plasma modelling has shown that
  inductive acceleration can convert Poynting flux directly into bulk
  kinetic energy in the relativistic flows driven by rotating
  magnetized neutron stars and black holes. Here, we generalize this
  approach by adding an ion fluid. Solutions are presented in which
  all particles are accelerated as the flow expands, with comparable
  power channeled into each of the plasma components. In an
  ion-dominated flow, each species reaches the limiting rigidity,
     according to Hillas' criterion, in a distance significantly
  shorter than in a lepton-dominated flow. These solutions support the
  hypothesis that newly born magnetars and pulsars are potential
  sources of ultra-high energy cosmic rays. The competing process of
  Poynting flux dissipation by magnetic reconnection is shown to be
  ineffective in low-density flows in which the conventionally defined
  electron multiplicity satisfies 
$\kappa_{\rm e}\lesssim  10^5\left(4\pi L_{38}/\Omega\right)^{1/4}
  /\textrm{Max}\left(\ionfraction^{1/2},1\right)$, where 
$L_{38}\times 10^{38}\textrm{erg s}^{-1}$ is the power carried by the flow in a
  solid angle $\Omega$, and $\ionfraction$ is the ratio of the ion to
  lepton power at launch.
\end{abstract}


\section{Introduction}
\label{intro}
Newly born magnetars, rapidly rotating pulsars and rotating black
holes, are thought to drive relativistic outflows that are dominated
by Poynting flux. They have long been recognized as promising
candidate sources of ultra-high energy cosmic rays \citep[for a review
see][]{koteraolinto11}, and have recently been suggested as sources of
PeV neutrinos \citep{icecube18}. However, the physics of the particle
acceleration processes at work remains uncertain
\citep[see, for example, the discussion in][]{lemoine13}, and most theoretical work has concentrated
on the problem of accelerating the leptons needed to explain the
highly variable gamma-ray emission that has been attributed 
to these objects.

For example, \lq\lq inductive acceleration\rq\rq, was 
advanced by \citet{kirkgiacinti17} 
as an explanation of the gamma-ray flares observed from the
Crab Nebula~\citep{buehlerblandford14}.  It involves
gradually extracting energy from the fluctuating component of the
magnetic field in a Poynting-flux dominated, electron-positron outflow
as it propagates radially, undisturbed by the surrounding
medium~\citep{kirkmochol11,kirkmochol11b}. In the case of the Crab,
acceleration by this process is limited because the pulsar wind terminates at
about $4\times10^{17}\,$cm from the pulsar, at which point only $10\%$
of the available energy has been channeled into relativistic
particles. Nevertheless, the mechanism injects leptons of several PeV
into the surrounding nebula.  

Motivated by the renewed interested in
the acceleration of hadrons in Poynting-flux dominated outflows
\citep[e.g.,][]{petropouloumastichiadis18,werneretal18,alvesetal18,reimeretal18,gaoetal19,zachariasetal19},
this paper generalizes inductive acceleration to 
outflows that contain a significant component of 
ions. We find that ions are indeed accelerated to high energy, that they 
are inevitably accompanied by a comparable flux of 
high energy leptons, and that they alleviate one of the main limitations of 
purely leptonic flows by substantially speeding up the acceleration process.

The fundamental assumption underlying inductive acceleration is that
the central object launches a relativistic, Poynting-flux dominated
flow, that can be considered to be radial and uniform within a given
solid angle. The analysis starts at a radius where the particle
density is high enough for the equations of ideal MHD to apply.
The flow is assumed to have negligible thermal pressure at this point, and to be causally detached from
the central object, in the sense that its velocity exceeds that of a
radially propagating fast magnetosonic wave. We show that the MHD description fails beyond a 
certain radius, 
because the plasma density decreases outwards. Short
length-scale variations in the toroidal magnetic field that are
frozen into the MHD flow, then gradually convert their Poynting flux
into radial bulk motion of the plasma. This continues until all
small-scale structure is erased, unless the flow previously
terminates by encountering an external medium. In
\S~\ref{poyntingoutflows} we motivate our assumptions and
discuss the nature of the frozen-in fluctuations. The fluid equations
that include ions are presented in \S~\ref{multifluid}.
Approximate analytic formulas describing the different phases of the flow 
are given and 
illustrated using numerical solutions of the
underlying equations
in \S~\ref{solutions}.
The physical interpretation of the solutions and their range of validity 
is discussed in more depth in 
\S~\ref{discussion}, where the mechanism is compared and contrasted with 
the related processes of reconnection and unipolar induction. Our conclusions 
are summarized in \S~\ref{conclusions}.

\section{Poynting flux dominated outflows}
\label{poyntingoutflows}

As is well-known \citep{buckley77}, the force-free MHD equations for
a steady, axisymmetric flow imply radial acceleration of its constituent charges such
that their Lorentz factor increases linearly with radius $r$ and the
magnetization parameter $\sigma$, which is the ratio of the radial
Poynting flux to the radial kinetic energy flux, decreases inversely
with $r$ \citep{contopoulosetal99,contopouloskazanas02,arons03}. 
\citet{bogovalov99} showed this property holds also for non-axisymmetric flows 
that contain
reversals of the magnetic field, provided these are 
concentrated in current sheets of thickness small compared to $r$. 
However, the force-free approximation neglects the inertia of the
particle component of the wind, which becomes increasingly important
towards larger radius, and the solution loses its validity when
the Lorentz factor of the flow approaches that of the fast
magnetosonic wave. Exactly where this happens depends on how particles are 
injected into the flow. In the idealized case of 
a cold axisymmetric wind, the fast magnetosonic point retreats to infinite radius
\citep{michel69,goldreichjulian70}. But, at least for pulsars, this is probably irrelevant, 
since models of pair production suggest particles are injected with Lorentz factors $\sim10^2$ to $10^3$, 
in which case the flow is 
supersonic already relatively close to the star \citep{lyubarskykirk01}, i.e., near 
the \lq\lq light cylinder\rq\rq\ at $r=r_{\rm L}$ ($=cP/2\pi$, where $P$ is the pulsar period).

Beyond the fast-magnetosonic point, acceleration must quickly
cease, since the plasma cools adiabatically, and 
a cold, radial, supersonic flow that obeys the ideal MHD
equations, propagates at constant
$\gamma$ and $\sigma$ (for a review, see \cite{kirketal09}).  In this
phase of the flow, the field reversals considered by
\citet{bogovalov99} become concentric current sheets separating regions
of toroidal magnetic field with opposite polarity --- a \lq\lq striped
wind\rq\rq. However, any fluctuations in the toroidal field that
appear in the co-moving frame to be static, local equilibria are
frozen into the flow and are convected radially with it.

As the plasma thins out further, a radius is reached at which 
also the ideal MHD approximation fails. For relatively high particle densities, 
this can happen if dissipation becomes important, for example, in the guise of reconnection
(see \citet{coroniti90}, \citet{lyubarskykirk01}, \citet{drenkhahnspruit02}, \citet{kirkskjaeraasen03},
and the discussion in \S~\ref{dissipation}). 
But, even without dissipation,  
the MHD approximation fails when the inertia
associated with the plasma current --- which is essential to preserve
the frozen-in fluctuations --- becomes important
\citep{usov75}. 

This effect was examined by \citet{kirkmochol11} (\lq\lq KM\rq\rq\ in the following), who
used a perturbation method to formulate equations governing the 
radial evolution of an electron-positron
wind with a frozen-in, sheared magnetic field, assuming each lepton species can be
described as a cold fluid. The main difference between this approach and  
those based on dissipation, is that the local equilibrium in the MHD flow is not 
assumed to be a pressure-supported 
structure, such as a hot current sheet, but is, instead, a force-free equilibrium. 
This relieves the problems associated with postulating that 
dissipation leads to an isotropic particle 
velocity distribution, which can be shown to be physically 
inconsistent beyond a critical, 
density-dependent, radius --- see \S~\ref{discussion} and \citet{lyubarskykirk01}. 

The particular force-free equilibrium chosen by KM
is a \lq\lq sheet-pinch\rq\rq\ \citep{bobrovaetal01,lietal03}, with
constant rate of shear.  Subsequently, 
a configuration with two field
reversals per wave period, concentrated in rotational
discontinuities,
was examined by 
\citet{kirkgiacinti17}. Here, we generalize two aspects of these
treatments.  Firstly, we show that the precise structure of the
equilibrium, in particular the thickness of the current-carrying layers, is not
important for the large-scale evolution of the flow. Secondly, and
more importantly, we examine the consequences of adding a fluid of
cold ions to the flow, assuming they do not contribute to the
transverse current and that their charge density is compensated by an
excess of electrons over positrons.

\section{The multi-fluid equations}
\label{multifluid}
We consider solutions of the multi-fluid equations that depend
spatially only on $r$, contain a transverse component of
the magnetic field that oscillates in the lab.\ frame with angular frequency $\omega
=c/r_{\rm L}\gg c/r$
and
are, to zeroth order in $r_{\rm L}/r$, stationary in a
frame comoving radially with the wave. Following KM, we denote the 
radial component of the four-speed (in units of $c$) of fluid $s$ 
($s=\,$e,i,p for electrons, ions and positrons, respectively) by 
${\ppars}$, and use the complex quantity $p_{\bot s}$ 
to denote the two components transverse to the radius vector. 
Since the radial component of the three-speed of each charged fluid
species equals the pattern speed $c\betaw$ of the wave, 
one has
\eqb
{\ppars}&=&\betaw\gamma_s,
\label{patternspeed}
\\
\noalign{\hbox{where}}
\gamma_s&=&\sqrt{1+{\ppars}^2+{\pperps}^2}
\nonumber\\
&=&\gammaw\sqrt{1+{\pperps}^2},
\label{frozen}
\eqe 
and $\gammaw=\left(1-\betaw^2\right)^{-1/2}$ is the Lorentz factor of the wave. 
KM~considered a purely leptonic
flow, in which case, the electron and positron flows can be assumed to
be symmetric in velocity: $\ppare=p_{\|{\rm p}}$ and 
$p_{\bot{\rm e}}=-p_{\bot{\rm p}}$, and have equal proper densities: 
$n_{\rm e}=n_{\rm p}$. Here we add a fluid containing ions of charge $Ze$, 
assuming (i) it does not
contribute to the transverse current, i.e., $p_{\bot{\rm i}}=0$, (ii) it does not
affect the symmetry of the electron-positron velocities, and (iii) its
charge density is compensated by an excess of electrons, such that the
flow remains neutral in the lab.\ frame: $Zn_{\rm i}=\gamma_{\rm e}
\left(n_{\rm e}-n_{\rm p}\right)/\gamma_{\rm i}$.  
As discussed in 
\ref{ampereslaw}, we assume $\pperps$ is independent of wave phase. It then follows 
from (\ref{patternspeed}) that $p_{\|{\rm s}}$ is also phase-independent, and 
the rate of shear is determined by the phase-dependence of the proper
densities.

\subsection{Conservation of particle number}
The continuity equation for the zeroth order quantities 
in the case of electron and positron fluids is given in KM, Eq.~(14), 
(henceforth, we refer to equations in KM as (KM~14), etc.). To account for
the presence of ions, we introduce the constant $\ionfraction$, which equals the ratio
of the rest-mass flux carried by ions to that carried by leptons. (At launch, $\ionfraction$
equals the ratio of the power carried by ions to that carried by leptons --- a quantity
that evolves as the flow accelerates.) The continuity 
equation (KM~14) then becomes:
\eqb
\left(1+\ionfraction\right)
\frac{r^2\ppare\omega_{\rm p}^2}{r_{\rm L}^2\omega^2}&=& a_{\rm Le}^2/\mu ,
\label{KM14}
\eqe
where the proper plasma frequency is defined as 
\eqb
\omega_{\rm p}&=&
\left[4\pi\left<n_{\rm e}+n_{\rm p}\right> e^2/m_{\rm e}\right]^{1/2},
\label{plasmafreq}
\eqe
and $\left<\dots\right>$ denotes a phase average. 
The right-hand side of Eq.~(\ref{KM14}) is $4\pi e^2/\left(m_{\rm e}^2c^3\right)$ times 
the phase-averaged 
flux of rest-mass per unit solid angle, $\diff \dot{M}/\diff\Omega$, 
expressed in terms of the following
dimensionless constants:
\begin{enumerate}
\item
The \lq\lq strength\rq\rq\ parameter
\eqb
a_{\rm Le}&=&
\left[4\pi\left(\diff L/\diff\Omega\right)e^2/m_{\rm e}^2c^5\right]^{1/2}
\nonumber\\
&=& 3.4\times 10^{10} \left(4\pi L_{38}/\Omega\right)^{1/2} ,
\nonumber
\eqe 
with $\diff L/\diff\Omega$ the power per unit solid angle carried by the flow, 
and $L_{38}$ the corresponding total power expressed in units of
$10^{38}\,\textrm{erg\,s}^{-1}$, assuming the flow occupies an effective
solid angle $\Omega$. 
In electron-positron plasmas, $a_{\rm Le}$ is a 
characteristic value of the maximum leptonic Lorentz factor. More generally, the 
parameter $a_{{\rm L}s}=\left(\left|q_s\right|m_{\rm e}/em_s\right)a_{{\rm Le}}$ gives the 
characteristic maximum Lorentz factor for particles of species $s$. 
In Poynting-flux dominated flows, the magnitude of the toroidal
magnetic field, $\left|B\right|$, is inversely proportional to $r$, and   
$a_{\rm Ls}$ equals the ratio of the gyro frequency of a (nonrelativistic) particle
of species $s$ to the wave frequency, when the fields 
are extrapolated back to radius $r_{\rm L}$:
\eqb
a_{\rm Ls}&=&\left|q_s B\right|r/\left(m_sc^2\right)
\nonumber\\
&=&\left|q_s B_{\rm L}\right|/\left(m_sc\omega\right)
\label{strengthparameter}
\eqe
\item
The inverse mass-loading:
\eqb
\mu&=&\left(\diff L/\diff\Omega\right)/\left(c^2\diff\dot{M}/\diff\Omega\right),
\eqe
where $\diff \dot{M}/\diff\Omega$ is the rest-mass flux per unit solid angle. 
In a pure lepton model, $\mu$ equals the Lorentz factor that would be achieved by the flow 
after conversion of the entire electromagnetic energy flux into kinetic energy.
\end{enumerate}
As an alternative to $\mu$, a 
more intuitive measure of the rest-mass flux is the particle multiplicity $\kappa_s$
at $r=r_{\rm L}$. This quantity is conventionally defined as the ratio of the 
lab.\ frame charge density to the \lq\lq Goldreich-Julian\rq\rq\ charge density 
$B_{\rm L}\omega/2\pi c$, where $B_{\rm L}$ is the magnitude of 
the toroidal magnetic field, extrapolated back to radius $r_{\rm L}$ \citep{lyubarskykirk01}.  
For an electron-positron plasma, $\kappa_{\rm e}=a_{\rm Le}/\left(4\mu\right)$ (see KM~7).
Adding an ion fluid, and introducing the notation $\kappaep$ for the 
sum of the electron and positron multiplicities leads to 
\eqb
\kappaep&=&\kappa_{\rm e}+\kappa_{\rm p}
\nonumber\\
&=&
a_{\rm Le}/\left[2\mu\left(1+\ionfraction\right)
\right].
\eqe
The ion multiplicity follows from the condition of charge
neutrality, 
$\kappa_{\rm i}=\kappa_{\rm e}-\kappa_{\rm p}$,
and can also be expressed in terms of $\ionfraction$:
\eqb
\kappa_{\rm i}&=&\kappa_{\rm ep}\ionfraction Z m_{\rm e}/m_{\rm i}.
\eqe
Equation~(\ref{KM14}), 
can then be reformulated using (\ref{patternspeed}) and (\ref{frozen}) 
to give an expression for the plasma frequency
as a function of $r$, $\pperp$ and $\gammaw$ (or, equivalently, $\betaw$):
\eqb
\omega_{\rm p}^2&=&
\frac{2\kappaep c^2 a_{\rm Le}}
{r^2\betaw\gammaw\sqrt{1+\pperp^2}},
\label{contfinal}
\eqe
where $\pperp=\left|p_{\bot{\rm e}}\right|=\left|p_{\bot{\rm p}}\right|$.

\subsection{Amp\`ere's Law}
\label{ampereslaw}

Seen from the comoving frame, the flow is locally
in force-free equilibrium, with the transverse
components of the electron and positron four-speeds, and, therefore, the 
plasma current, directed along the local magnetic field.
Then, in the absence of plasma
pressure, the (complex) 
transverse magnetic field $B$ simply rotates as a function of wave phase, 
$\varphi$,  
keeping its magnitude constant to zeroth order. 
This behavior follows from Amp\`ere's law, which 
links the (complex) transverse current density $j_\bot$ with $B$ according to (KM~13):
\eqb
\frac{\partial B}{\partial\varphi}
&=&
\frac{4\pi \imath j_\bot \betaw\gammaw^2}{\omega}
\nonumber\\
&=&
\frac{4\pi \imath \left(p_{\bot{\rm p}} n_{\rm p} - p_{\bot{\rm e}}n_{\rm e}\right)
ec \betaw\gammaw^2}{\omega}.
\eqe
Integrating this equation over a phase interval in which $B$ reverses its sign
one finds
\eqb 
\int_{\rm
reversal}\diff\varphi \left|j_\bot\right|&=&\left(\left|B\right|/\gammaw\right)/
\left(4\betaw\gammaw/\omega\right).
\label{jperp2}
\eqe 
(Note that $\left|B\right|/\gammaw$ is the amplitude of the magnetic field in the
frame co-moving with the wave, and $2\pi c\betaw\gammaw/\omega$ is the
wavelength in this frame.)

Equation (\ref{jperp2}) allows some freedom in specifying 
the phase-dependencies of the transverse fluid velocities and densities. 
For simplicity, we treat here the case where 
$\left|p_{\bot {\rm e}}\right|=\left|p_{\bot {\rm p}}\right|$ is constant, and the entire
phase-dependence of the amplitude of the current arises from that of the density. Then, 
assuming two, not necessarily thin, field reversals per wave period, 
the averaged proper particle densities 
$\left<n_{\rm e,p}\right>=\int_0^{2\pi}n_s\diff\varphi/\left(2\pi\right)$ obey: 
\eqb
\int_{\rm reversal} \diff\varphi \left|j_\bot\right|
&=& ecp_\bot\pi \left<n_{\rm e} + n_{\rm p}\right>
\eqe
so that, from (\ref{jperp2}), and (\ref{plasmafreq}),
\eqb
\left|B\right|&=&\frac{m_{\rm e}c\betaw\gammaw^2\pperp\omega_{\rm p}^2}{e\omega}.
\label{ampere}
\eqe
Eliminating $\omega_{\rm p}$ using the continuity equation (\ref{contfinal}) gives:
\eqb
\left|B\right|&=&\frac{m_{\rm e}c^3 2\kappaep a_{\rm Le}\pperp\gammaw}{e\omega 
r^2\sqrt{1+\pperp^2}}.
\label{amperefinal}
\eqe

\subsection{Conservation of energy}
The equation of conservation of energy can be written as 
\eqb
\mu&=&\Psi_{\rm Poynting}  +  \Psi_{\rm particle}
\label{energyfinal}
\eqe
where, $\Psi_{\rm Poynting}$ is the normalized Poynting flux:
\eqb
\Psi_{\rm Poynting}&=&\frac{\betaw r^2 \left|B\right|^2}
{4\pi c\left(\diff\dot{M}/\diff\Omega\right)}
\eqe
and $\Psi_{\rm particle}$ is the normalized kinetic 
energy flux:
\eqb
\Psi_{\rm particle}&=& r^2\sum m_s c \left<n_s\right>\gamma_s\ppars/
\left(\diff \dot{M}/\diff\Omega\right).
\eqe
(see (KM~A26)). 
The purely leptonic case is  
particularly simple, 
since the energy flux carried by particles is just 
$\gamma_{\rm e} c^2$ times the particle rest-mass flux.
In terms of the 
magnetization parameter, $\sigma=\Psi_{\rm Poynting}/\Psi_{\rm particle}$, one then finds
(KM~15) $\mu=\gamma_{\rm e}\left(1+\sigma\right)$. In the presence of ions, 
however, this equation 
is modified because the ion energy flux is 
$\gammaw c^2$ times the ion rest-mass flux, whereas the leptonic energy flux is 
$\gamma_{\rm e} c^2$ times the leptonic rest-mass flux, so that 
\eqb
\Psi_{\rm Poynting}&=&\frac{\mu\betaw}
{a_{\rm Le}^2}
\left(\frac{er\left|B\right|}{m_{\rm e}c^2}\right)^2
\nonumber\\
&=&\frac{2\kappaep a_{\rm Le}r_{\rm L}^2}{\left(1+\ionfraction\right)}
\left(\frac{\betaw\gammaw^2\pperp^2}{r^2\left(1+\pperp^2\right)}\right)
\label{poyntingfinal}
\\
\noalign{\hbox{and}}
\Psi_{\rm particle}&=&
\left(\frac{\ionfraction+\sqrt{1+p_\bot^2}}{1+\ionfraction}\right)\gammaw .
\label{particlefinal}
\eqe
where, in (\ref{poyntingfinal}), Amp\`ere's equation~(\ref{amperefinal}) 
was used to eliminate $\left|B\right|$. With these definitions, (\ref{energyfinal}) is 
an algebraic equation relating the $r$-dependent variables $\pperp$ and $\gammaw$. 

\subsection{Radial momentum balance equation}
The radial momentum flux per unit mass, $\nu$, is similar in form to the 
normalized energy flux $\mu$
in Eq.~(\ref{energyfinal}):
\eqb
\nu
&=&\Pi_{\rm Poynting}+\Pi_{\rm particle}
\label{normalizedmomentum}
\eqe
where $\Pi_{\rm Poynting}$ is the normalized electromagnetic momentum flux per unit solid angle:
\eqb
\Pi_{\rm Poynting}
&=&\left(\frac{1+\betaw^2}{2\betaw}\right)\Psi_{\rm Poynting}
\eqe
and $\Pi_{\rm particle}$ is the particle contribution:
\eqb
\Pi_{\rm particle}
&=&\betaw\Psi_{\rm particle}
\eqe
However, radial momentum is not conserved. 
In the purely leptonic case, this leads to equation (KM~16), 
the right-hand side of which is modified by the presence of ions 
(cf.~KM~A27) to give:
\eqb
\frac{\diff\nu}{\diff r}&=&\frac{\mu r \pperp^2\omega_{\rm p}^2}
{a_{\rm Le}^2} 
\nonumber\\
&=&\frac{\pperp^2}{\left(1+\ionfraction\right) \betaw\gammaw r\sqrt{1+\pperp^2}}
\label{radialmomentum}
\eqe

In principle, this first-order,  
ordinary differential equation for the zeroth-order functions $\pperp$ and $\gammaw$ 
closes the system, when combined with the algebraic constraint (\ref{energyfinal})
and the definitions (\ref{poyntingfinal}) and (\ref{particlefinal}). 
However, for a relativistic flow, $\Psi_{\rm Poynting}\approx\Pi_{\rm Poynting}$ and
$\Psi_{\rm particle}\approx\Pi_{\rm particle}$. 
Because of this, it is preferable to
reformulate (\ref{radialmomentum})
by subtracting the conserved energy flux 
from the momentum flux (times $c$).
Expanding $\nu$ and $\mu$ in $1/\gammaw$, 
one then arrives at the radial momentum balance equation:
\eqb
\frac{\diff}{\diff r}\left[
\frac{\kappaep a_{\rm Le} r_{\rm L}^2\pperp^2}
{4r^2\gammaw^2\left(1+\pperp^2\right)}
-
\frac{\ionfraction+\sqrt{1+\pperp^2}}{2\gammaw}
\right]&&
\nonumber\\
\,=\,\frac{p_\bot^2}{r\gammaw\sqrt{1+p_\bot^2}},&&
\label{reducedradialmomentum}
\eqe
which is used in place of (\ref{radialmomentum}) to determine the solutions described below.

\begin{figure*}
\centering
\input{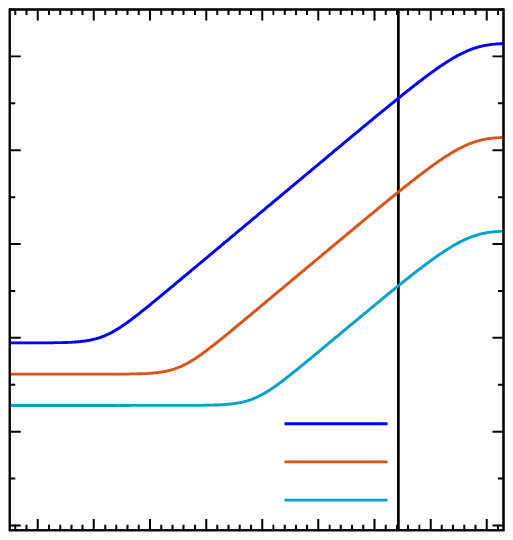}\input{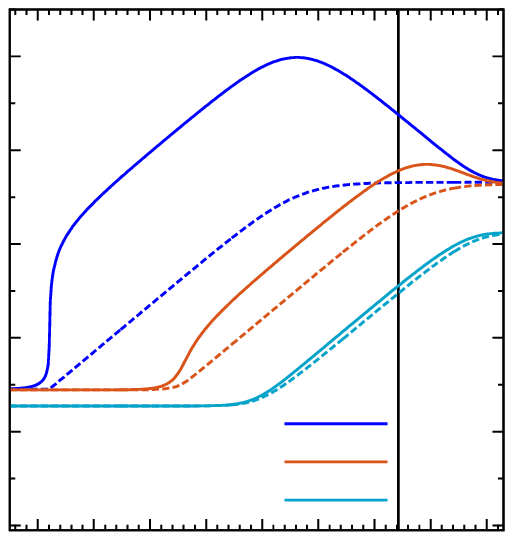}
\caption{
Numerical integration of the multi-fluid equations for a pure electron-positron plasma,
with $\kappa_{\rm i}=0$, (left-hand panel), and 
for a plasma containing also a proton fluid 
with $\kappa_{\rm i}=1$, (right-hand panel), for electron 
multiplicities $\kappa_{\rm e}=1$, $10^2$, and $10^4$, corresponding, in the right-hand panel, 
to $\ionfraction=1836$, $9.2$ and $0.09$, respectively.
In each case the strength parameter
$a_{\rm Le}=7.6\times10^{10}$, appropriate for the wind of the Crab pulsar, and
the flow is launched at Mach number ${\cal M}=5$. The vertical black line indicates the location 
of the termination shock in the Crab Nebula (in which $r_{\rm L}=1.6\times10^8\,$cm).
}
\label{figone}       
\end{figure*}

\section{Properties of the solutions}
\label{solutions}
It is a good
  approximation to set $\betaw=1$ when inserting expression
  (\ref{poyntingfinal}) for the Poynting flux into the energy equation 
(\ref{energyfinal}), since the flows we
  consider are everywhere relativistic. This results in a quadratic
  equation, which can be solved to give $\gammaw$ as a
  function of $r$ and $\pperp$.  The radial momentum balance equation
  (\ref{reducedradialmomentum}) then reduces to a first-order ordinary
  differential equation for $\pperp$ as a function of $r$, which is
  easily integrated using a standard algorithm.  However, several key
  properties of the outflow can be deduced purely analytically.  

If Poynting flux dominates and the flow is relativistic, it follows from 
energy conservation (\ref{energyfinal}) and (\ref{poyntingfinal}) that 
\eqb
\gammaw&\approx&\frac{r\sqrt{1+\pperp^2}}
{2\kappaep r_{\rm L} \pperp}.
\label{Pdominatedphase}
\eqe 
Therefore, in the MHD phase, where $p_\bot\ll1$ and
$\gammaw$ is constant, one has $p_\bot\propto r$, which demonstrates explicitly 
that the MHD approximation 
fails at a critical radius.
Furthermore, beyond this radius, where $p_\bot\gtrsim 1$, 
the flow must enter an acceleration phase with
$\gammaw\propto r$, which persists until the kinetic energy flux 
becomes comparable in magnitude to the Poynting flux. 
This conclusion holds for both purely leptonic flows and for flows containing ions.

The critical radius, $r_{\rm MHD}$,  at which MHD fails
can then be estimated from (\ref{Pdominatedphase}) to be
\eqb
r_{\rm MHD}&\approx&\sqrt{2}r_{\rm L}\kappaep\gamma_{{\rm w}0}
\nonumber\\
&=&2^{1/6}r_{\rm L}\left(\frac{
a_{\rm Le}{\cal M}^2\kappaep^2}
{1+\ionfraction}\right)^{1/3}
\nonumber
\\
&=&3.6\times 10^3 
r_{\rm L} 
\left(\frac{{\cal M}^2\kappaep^2}
{1+\ionfraction}\right)^{1/3}
\left(\frac{4\pi L_{38}}{\Omega}\right)^{1/6} 
\label{rmhd}
\eqe 
where ${\cal M}=\gamma_{{\rm w}0}/\sigma_0^{1/2}
=\gamma_{{\rm w}0}^{3/2}/\mu^{1/2}\gtrsim 1$ is the
relativistic Mach number of the flow at launch, at which point $\gammaw=\gamma_{{\rm w}0}$
and $\sigma=\sigma_0$.

The properties of these flows are illustrated by the examples presented 
in Fig.~\ref{figone}, which depicts
numerical solutions, obtained using {\em Mathematica}, 
of Eqs.~(\ref{reducedradialmomentum}) and (\ref{energyfinal}), 
for
both a pure electron-positron plasma (left-hand panel) and a plasma containing also
protons (right-hand panel), assuming that the two current layers are
symmetrically located in wave phase, i.e., that the phase-averaged
magnetic field vanishes. (It is straightforward to lift this
assumption, see \citet{kirkgiacinti17}.)  In each case, three
regimes can be identified: (i) a mildly supersonic MHD flow
extending from the launching point to $r\approx r_{\rm MHD}$, 
in which the Lorentz factors of the fluids are approximately 
constant and 
equal to that of the pattern $\gammaw$, 
(ii) an acceleration phase at $r_{\rm MHD}<r<r_{\rm max}$, 
(see Eq.~(\ref{rmaxdef})) during which $\gammaw\propto r$, 
and the Poynting flux steadily decreases, and (iii) a
coasting phase at $r>r_{\rm max}$, where the plasma
relaxes to an outflow with all species moving radially at the same,
constant Lorentz factor, and vanishingly small Poynting flux.  

The main feature that distinguishes ion-carrying 
flows from pure electron-positron flows is the rapid energization of the
leptons at the beginning of the acceleration phase. The transverse velocity of the 
leptons initially rises, and then stabilizes at a value
$p_{\bot{\rm eq}}>1$ that remains constant until the end of the acceleration phase
at $r\approx r_{\rm max}$. 
This arises
because the radial acceleration of the ion fluid must
equal that of the lepton fluids. In the case of the ions, radial
acceleration is provided by the phase-averaged effect of the
first-order, radial electric fields, which, since the plasma is neutral, extract
the same momentum from the lepton fluids. However, the leptons over-compensate by acquiring
significant toroidal momentum, thereby enhancing the 
$\vec{j}_\bot\wedge \vec{B}$ and the centrifugal
terms in their radial equation of motion, both of which are of first order, since the 
current is closely aligned to the magnetic field, and the centrifugal force is 
inversely proportional to the radius.

In both the MHD and the acceleration zones, Poynting flux
dominates the flow, and one can use (\ref{Pdominatedphase}) to simplify the 
radial momentum balance equation (\ref{reducedradialmomentum}):
\eqb
\frac{\diff\pperp}{\diff r}&=&
\frac{\pperp\left(1+\pperp^2\right)}{r}\times
\nonumber\\
&&\left[\frac{1-\pperp^2+\ionfraction\sqrt{1+\pperp^2}-
\frac{4\kappaep^2a_{\rm Le}r_{\rm L}^3\pperp^3}{r^3\left(1+\pperp^2\right)}}
{\left(1+\pperp^2\right)^2+\ionfraction\sqrt{1+\pperp^2}-
\frac{4\kappaep^2a_{\rm Le}r_{\rm L}^3\pperp^3}{r^3\left(1+\pperp^2\right)}}
\right].
\label{dpperpdr}
\eqe
In the acceleration zone, the terms arising from Poynting flux 
(those containing $a_{\rm Le}$) are negligible, and one finds that  
$\pperp$ saturates at
\eqb
\pperpeq
&=&\left(\ionfraction^2+\ionfraction\sqrt{\ionfraction^2+8}+2\right)^{1/2}/\sqrt{2}.
\label{usaturation}
\eqe
I.e., 
$\pperpeq \approx1$ in a pair-dominated flow ($\ionfraction\ll 1$) and 
$\pperpeq\approx\ionfraction$ in an ion dominated flow ($\ionfraction\gg1$). 
Since the lepton Lorentz
factors in the acceleration phase are 
$\gamma_{\rm e}=\gamma_{\rm p}=\gammaw\sqrt{1+\pperp^2}$, a 
flow in which ions dominate the rest-mass flux, achieves
equipartition between the energy fluxes carried by ions and 
leptons in the acceleration phase.  

Another feature of flows containing ions is that the radius at which acceleration sets in 
decreases as $\ionfraction$ increases, as indicated by Eq.~(\ref{rmhd}).
This, together with the rapid energization described above, causes the maximum Lorentz factor 
to be reached much sooner than in purely leptonic flows. 
For example, Fig~\ref{figone} shows that when $\kappa_{\rm i}=1$, $\kappa_{\rm e}=1$ and 
$\ionfraction=1836$ (corresponding to protons), the 
acceleration starts at roughly $4\times 10^3r_{\rm L}$ and ends at 
$10^8 r_{\rm L}$ --- in the case of the Crab, 
well before the termination shock is reached. 
On the other hand, for protons with $\kappa_{\rm i}=1$, in a wind with $\kappa_{\rm e}=10^4$,
there is essentially no deviation from the corresponding case with a pure 
electron-positron plasma, shown in the left-hand panel. 
Such a lepton
dominated outflow with $\kappa_{\rm e}=1$ starts to accelerate at $5\times 10^4r_{\rm L}$ 
in the case of the Crab,
and converts only 10\% of its Poynting flux into kinetic energy before terminating. 

In the acceleration phase, where $p_{\perp}\approx p_{\perp{\rm eq}}$, one finds
from Eqs.~(\ref{frozen}) and (\ref{Pdominatedphase})
\eqb  
\gamma_{\rm i}&\approx&r/\left(2\kappaep r_{\rm L}\right)
\nonumber\\
\gamma_{\rm e}&\approx&\ionfraction\gamma_{\rm i},
\label{grion}
\eqe
for the ion dominated case ($\ionfraction\gg1$), and 
\eqb
\gamma_{\rm e}&\approx&r/\left(\kappaep r_{\rm L}\right),
\label{grlepton}
\eqe  
for the lepton dominated case, ($\ionfraction\ll1$).

At the end of the acceleration phase, the Poynting flux is negligible and
the maximum Lorentz factors achieved are 
given by (\ref{energyfinal}), setting $\Psi_{\rm Poynting}\rightarrow0$, leading to
\eqb
\gamma_{\rm i, max}&\approx&a_{\rm Li}/\left(4\kappa_{\rm i}\right)
\nonumber\\
\gamma_{\rm e,max}&\approx&a_{\rm Le}/\left(4\kappaep\right)
\label{maxenergy}
\eqe
for $\ionfraction\gg1$
and
\eqb
\gamma_{\rm i,max}&=&a_{\rm Li}/\left(2\sqrt{2}\kappa_{\rm i}\right)
\nonumber\\
\gamma_{\rm e,max}&=&a_{\rm Le}/\left(2\kappaep\right)
\eqe
for $\ionfraction\ll1$.
These estimates hold
provided $\kappa_{\rm i}>1$ and $\kappaep>1$ --- see the discussion of 
the range of validity of the fluid approximation in \S~\ref{discussion}.
The radius at the end of the acceleration zone follows from 
(\ref{grion}) and (\ref{grlepton}):
\eqb
r_{\rm max}\approx
a_{\rm Le}r_{\rm L}/2\left(1+\ionfraction\right),
\label{rmaxdef}
\eqe
showing that conversion of electromagnetic energy into particle energy takes place
much more rapidly when ions dominate. 

For $r>r_{\rm max}$, the flow enters a coasting phase, in which,
according to the equations derived here, the
leptons relax to the same Lorentz factor as the ions, and the flow
subsequently proceeds at constant velocity in the absence of the wave
component of the fields. However, as we discuss below, the validity of the perturbation
expansion underlying this description is doubtful when $r\gtrsim\kappa_{\rm e,i}r_{\rm max}$.

\begin{figure*}
\centering
\input{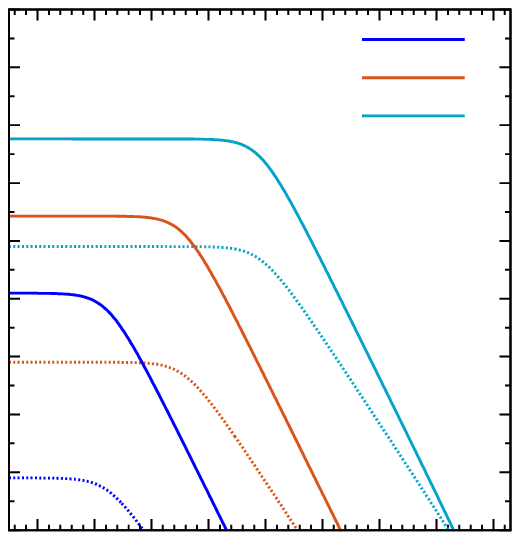}\input{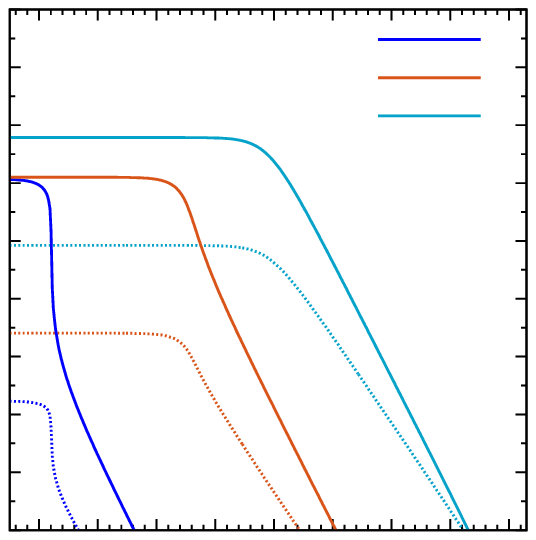}
\caption{Ratios of the proper plasma frequency $\omega_{\rm p}$ (dotted lines)
and the electron gyro frequency 
$\omega_{\rm g}$ (solid lines)
to the dynamical expansion rate $\omega_{\rm dyn}$ (see 
Eqs.~(\ref{plasmaratio}) and (\ref{gyroratio})) 
for the solutions presented in Fig.~\ref{figone}. 
}
\label{figtwo}       
\end{figure*}

\section{Discussion}
\label{discussion}

The solutions presented in \S~\ref{solutions} are based on a
perturbation analysis of the multi-fluid equations, using the small
parameter $r_{\rm L}/r$.  This results in a simple, closed system of
equations describing the radial evolution of the zeroth-order
quantities in a quasi-stationary, Poynting-flux dominated flow, once
it has left the vicinity of the rotating, magnetized object that
launches it. The nature of the first-order fields plays no role,
provided only that they remain small compared to the zeroth-order
fields. They may, for example, be rapidly oscillating functions of
wave phase. In principle, solutions could be found analytically for
the these quantities.  However, even within this restricted, multi-fluid
description, the number of degrees of freedom available to the plasma
is considerable, and the choice of initial conditions largely
unconstrained, so that it would be difficult to extract additional
insights from particular, first-order solutions.

Alternatively, the multi-fluid equations could be solved numerically
without recourse to perturbation theory. Several groups have developed
appropriate algorithms, and applied them to special situations, such
as a reconnecting current sheet, or a Poynting-flux dominated
termination shock
\citep{zenitanietal09,kojimaoogi09,amanokirk13,barkovkomissarov16,amano16}. The
results reveal structures similar to those observed in particle-in-cell simulations, 
which, at least in principle, incorporate more of the microphysics. However, 
the multi-fluid algorithms introduce arbitrary dissipative terms such as viscosity and 
resistivity, that are absent in the model with cold fluids presented here, and that might
mask the importance of the inertial effects responsible for inductive acceleration.

In the perturbative approach, the zeroth-order electric field is
orthogonal to the zeroth-order fluid velocity, and acceleration is the
result of higher order effects associated with the motion of
cold fluids, which, to this order, are no longer tied to magnetic
field lines. This picture fails at the microscopic level 
if the fluids do not remain cold, or if
the flow contains too few particles for them to be represented as a
fluids. In the following we discuss these limitations in turn.

\subsection{Dissipation and magnetic reconnection}
\label{dissipation}
The cold fluid assumption breaks down if instabilities result in the 
dissipation a significant amount of energy into heat. 
The sheet pinch equilibrium assumed in \S~\ref{multifluid} is stable within the 
ideal MHD description. Furthermore, it is not subject to the 
kinetic tearing mode instability, since the corresponding growth rate 
vanishes for a cold plasma. Nevertheless, counter-streaming electron and positron 
fluids are subject to a Buneman-type electrostatic instability 
\citep{lietal03}, which leads to randomization of the 
component of the leptonic velocity along the local magnetic field. The instability
saturates when the 
velocity dispersion reaches the drift speed $cp_\bot/\gamma_{\rm e}$, 
which is small in the MHD regime, 
but could, in principle, lead to an anomalous resistivity, and the destabilization 
of the tearing mode \citep{komissarovetal07}. In this case
the subsequent nonlinear evolution generates
a turbulent magnetic field component of the same order of magnitude as the 
regular component, as 
confirmed by numerous simulations \citep[for a recent review, see][]{kaganetal15}.

The growth rate of the electrostatic instability, 
seen in the frame co-moving with the wave
(the \lq\lq wave frame\rq\rq), 
is of the order of the proper plasma frequency, $\omega_{\rm p}$.
Assuming this growth is sufficiently rapid, the subsequent reconnection 
rate is limited from above  
by the rate at which the particle 
distribution can be isotropized, i.e., by the gyro-frequency $\omega_{\rm g}$ 
of an electron of four-velocity $p_\bot$ moving in the magnetic field $\left|B\right|/\gammaw$
seen in the wave frame:
\eqb
\omega_{\rm g}&=&
\frac{e \left|B\right|}{\gammaw\sqrt{1+p_\bot^2}m_{\rm e}c}
\nonumber\\
&=&\frac{\omega a_{\rm Le}r_{\rm L}}{\gamma_{\rm e}r}.
\label{gyrofreq}
\eqe
To assess whether or not dissipation is possible, 
these rates must be compared to the dynamical rate of evolution imposed by 
expansion of the flow 
\eqb
\omega_{\rm dyn}&=&\gammaw c/r ,
\label{expansionrate}
\eqe
which is 
measured in the wave frame at fixed phase. 
Using 
Eq.~(\ref{contfinal}), 
one finds
\eqb
\frac{\omega_{\rm p}}{\omega_{\rm dyn}}&\approx&
\frac{\left(2 a_{\rm Le}\kappaep\right)^{1/2}}
{\gammaw^{3/2}\left(1+p_\bot^2\right)^{1/4}}
\label{plasmaratio}\\
\noalign{\hbox{and}}
\frac{\omega_{\rm g}}{\omega_{\rm dyn}}&\approx&
\frac{a_{\rm Le}}
{\gammaw^{2}\left(1+p_\bot^2\right)^{1/2}}.
\label{gyroratio}
\eqe

Figure~\ref{figtwo} depicts these ratios for the illustrative cases discussed in 
\S~\ref{solutions} and shown in Fig.~\ref{figone}. 
During the MHD phase, both $\omega_{\rm p}/\omega_{\rm dyn}$ and
$\omega_{\rm g}/\omega_{\rm dyn}$ remain constant and greater than
unity for all cases studied except for the low density, pair plasma
case $\kappa_{\rm i}=0$, $\kappa_{\rm e}=1$, where 
$\omega_{\rm p}/\omega_{\rm dyn}\approx 1$.  
This does not establish the importance of dissipation in the MHD phase, but 
it also does not rule it out.
More importantly, the relevance of 
microphysical processes declines rapidly when the
acceleration phase is entered, and both rates drop below the dynamical
expansion rate at some point before acceleration is complete, provided
the lepton density in the wind is sufficiently low. Inserting the expressions for 
$\gamma_{\rm max}$, Eq.~(\ref{maxenergy}) into (\ref{plasmaratio}) and (\ref{gyroratio})
one finds that $\omega_{\rm p}$ and $\omega_{\rm g}$ drop below the expansion rate
before the end of the acceleration phase if  
\eqb
\kappa_{\rm e}&\lesssim& \sqrt{a_{\rm Le}/\textrm{Max}\left(\ionfraction,1\right)}
\nonumber\\
&\approx& 10^5 \left(4\pi
  L_{38}/\Omega\right)^{1/4}/\textrm{Max}
\left(\ionfraction^{1/2},1\right),
\label{kappalimit}
\eqe
which confirms 
the point previously made by \citet[][see their Eq.~(37)]{lyubarskykirk01}. 

Several groups have analyzed Poynting-flux dominated flows assuming,
on the contrary, that dissipation by reconnection is important
\citep{lyubarskykirk01,drenkhahnspruit02,drenkhahn02,kirkskjaeraasen03,lyubarsky10,begue17,gianniosuzdensky19}. 
The existence of a hot current sheet that can
be described by an isotropic pressure tensor is implicit
in all of these investigations, but different prescriptions are used
to specify the dissipation rate.  In each case, 
bulk acceleration of the flow is found: $\gammaw\propto r^{q}$, typically with
$q\sim 1/3$ to $1/2$. 
However, for low density winds, Eq.~(\ref{kappalimit}) shows that 
acceleration inevitably leads into a
regime in which dissipation is frozen out 
before it is complete, and it is reasonable to expect that the evolution subsequently proceeds as
in the inductive solutions presented in \S~\ref{solutions}. 

Without making assumptions about the dissipation and isotropization
rates, several sets of PIC simulations have demonstrated
that magnetic structures frozen into a Poynting-flux dominated wind
do indeed dissipate
\citep{petrietal15,zrake16,zrakearons17,alvesetal18}.
However, these simulations are limited to small spatial regions, in which the
expansion of the flow is neglected. They are, therefore, relevant only to high density flows
which are able to dissipate a substantial fraction of their
electromagnetic energy before the microphysical processes 
freeze out.
Global simulations \citep{ceruttiphilippov17} offer a more generally applicable
approach, but
they are expensive in terms of computing resources. To date, they have 
been used to study the launch of only relatively weak 
waves, and do not yet extend into the acceleration phase predicted in \S~\ref{solutions}. 

\subsection{The fluid approximation and the Hillas Limit}

Assuming an undisturbed flow, the maximum Lorentz factor achieved in
the current model, given by Eq.~(\ref{maxenergy}), is inversely
proportional to the multiplicities.  Since a fluid
description is used, this expression must break down at sufficiently
small multiplicity, when there are too few particles for the star to launch a 
wind that can be described by the MHD equations. 
Exactly when this happens is unclear. One plausible constraint 
for ensuring fluid behavior is $\kappa_{\rm e,i}\gtrsim 1$, since, when this is violated,
superluminal waves are able to propagate at $r>r_{\rm L}$ \citep{mocholkirk13} and could
be launched in place of the MHD flow.  As a consistency check on this constraint, 
the
magnitude of the phase-averaged, first-order effects
responsible for acceleration can be estimated by ascribing them to an effective 
radial electric field $E^{(1)}_{\rm e,i}$, defined as 
\eqb 
E^{(1)}_{\rm e,i}&=&\frac{m_{\rm e,i}c^2}{\left|q_{\rm e,i}\right|}
\frac{\diff\gamma_{\rm e,i}}{\diff r}.
\label{effectiveEzero}
\eqe
Then, from Eq.~(\ref{Pdominatedphase}), noting that $p_\bot$ is constant during the acceleration phase, 
\eqb
E^{(1)}_{\rm e,i}/\left|B\right|&\approx&r/\left(r_{\rm max}\kappa_{\rm e,i}\right).
\label{effectiveE}
\eqe 
Therefore, the perturbation approach loses validity when
when $r$ approaches $\kappa_{\rm e,i}r_{\rm max}$ and 
the choice $\kappa_{\rm e,i}>1$ ensures that 
this does not happen until the flow has entered the final coasting phase.

It is interesting to compare the limit implied jointly by Eq.~(\ref{maxenergy}), together with
the choice $\kappa_{\rm e,i}>1$, with the limit on rigidity, $R_{\rm H}$,
given by \citet{hillas84}: $R_{\rm H}<(v/c)^2 \bar{B}\bar{r}$, where $\bar{r}$ and $\bar{B}$
are a characteristic length and magnetic field strength, and $v$ is a bulk flow velocity.  These limits 
coincide if $v$ is assumed to be close to $c$, and
$\bar{B}$ is interpreted as the amplitude $\left|B\right|$ 
of the oscillating, zeroth-order
field in the wind at radius $r=\bar{r}$.  In our case, the limit
is not strict, since it is imposed by the range of
validity of the approximations employed. In particular, our treatment
formally leaves open the possibility that instabilities in flows with
$\kappa_{\rm e,i}<1$ might accelerate particles to higher rigidity. However, to
do this, the flow must generate fields at large radius whose
strength exceeds that of the decaying wave component, placing it out of reach of 
our perturbative description. Of course, if $\bar{B}$ is reinterpreted in terms of these 
amplified fields, the Hillas limit remains valid.

\subsection{Radiation losses}
\label{radiationlosses}
To zeroth order in the small parameter $r_{\rm L}/r$, the acceleration suffered by the charged fluids in these solutions
vanishes. Furthermore,
first-order acceleration in the effective, phase-averaged electric field given by (\ref{effectiveE}) 
is parallel to the zeroth-order fluid velocity, which severely
reduces the radiation losses. Nevertheless, it is reasonable to expect that 
phase-dependent components of the electromagnetic fields perpendicular to the fluid velocity 
might also exist and be of similar magnitude. In this case, a rough estimate of the energy loss rate by radiation
of the lepton fluids can be found from Larmor's formula:
\eqb
\omega_{\rm loss}&=&\frac{1}{\gamma_{\rm e}}\frac{\diff\gamma_{\rm e}}{\diff t}
\nonumber\\
&=&\frac{2e^4}{3m_{\rm e}^3c^5}\gamma_{\rm e}\left(E^{(1)}_{\rm e}\right)^2
\eqe
In most cases, the corresponding radiation process is synchrotron radiation, although it should be
noted that Larmor's formula is quite general, remaining valid even for motion in rapidly oscillating 
electromagnetic fields, a process sometimes referred to as {\em jitter} radiation. 
Inserting the estimate given in (\ref{effectiveEzero}) and dividing the resulting loss rate
by the dynamical expansion rate gives
\eqb
\frac{\omega_{\rm loss}}{\omega_{\rm dyn}}&=&
\left(\frac{2e^2\omega}{3mc^3}\right)\left(\frac{r_{\rm L}\gamma_{\rm e}^2
\sqrt{1+\pperp^2}}{r}\right)
\left(\frac{r}{\gamma_{\rm e}}\frac{\diff \gamma_{\rm e}}{\diff r}\right)^2
\eqe
In the acceleration region, where $\gamma_{\rm e}\propto r$, losses become progressively 
more important at larger radius, and, at $r\approx r_{\rm max}$, one has, to order of magnitude,
\eqb
\frac{\omega_{\rm loss}}{\omega_{\rm dyn}}&\approx&
\frac{2e^4}{3m_{\rm e}^3c^5}\textrm{Max}\left(1,\ionfraction^2\right) a_{\rm e}
\nonumber\\
&\approx& 10^{-13}\textrm{Max}\left(1,\ionfraction^2\right) L_{38}^{1/2}P_{\rm sec}^{-1},
\eqe
where $P_{\rm sec}$ is the wave period expressed in seconds. Thus, radiation losses in the acceleration zone are negligible for pulsars, 
but could enter into play in the distant regions of a flow driven by 
a protomagnetar, where $L_{38}\approx10^{16}$ and $P_{\rm sec}\approx 10^{-3}$ \citep{metzgeretal11}.

  In the inner parts of the flow, the transition from the MHD to the
  acceleration zone produces rapid energization of the leptons, in
  particular if ions dominate the rest-mass flux. From
  (\ref{dpperpdr}) one can estimate that, for $\ionfraction\gg1$,
  $\diff\ln\gamma_{\rm e}/\diff\ln r\sim\ionfraction/\pperp$ in the zone
  close to $r=r_{\rm MHD}$, and, therefore, 
\eqb 
\frac{\omega_{\rm
      loss}}{\omega_{\rm dyn}}&\approx&
  \left(\frac{2e^2\omega}{3mc^3}\right)
\ionfraction^{8/3}\kappaep^{-4/3}a_{\rm Le}^{1/3}{\cal M}^{2/3}
\nonumber\\
&\approx&10^{-19}
\ionfraction^{8/3}\kappaep^{-4/3}{\cal M}^{2/3}
L_{38}^{1/6}P_{\rm sec}^{-1}.
\eqe
Radiation losses in this part of the flow are, therefore, dynamically unimportant, unless
the wind is launched at a very high Mach number.
However, this statement applies only to losses caused by the electromagnetic fields 
carried by the wind.
Other mechanisms, such as inverse Compton scattering off the ambient photon field, 
could, in principle, be important in objects such as protomagnetars, blazars and 
in the phenomena powering gamma-ray bursts.

\subsection{Migration in latitude and the unipolar inductor}
In addition to the limitations placed on the model by microphysical effects, 
another potential 
limitation arises at the global level, because of the assumption that the flow
is uniform and radial within a solid angle $\Omega$. 
Since $p_{\bot s}/p_{\| s}\le \left(\betaw\gammaw\right)^{-1}\ll 1$, the fluid trajectories indeed remain radial to a good approximation. 
However, as an individual fluid element moves outwards, its radius vector drifts slowly in latitude and/or longitude --- depending on the wave phase at which it is located --- 
implying that the assumption of uniformity within $\Omega$ is valid only if the angular displacement $\Delta\theta$ accumulated by each element 
between launch and $r=r_{\rm max}$ 
satisfies $\Delta\theta<\Omega^{1/2}$.  
The rate at which the radius vector rotates is given by
\eqb
\frac{\diff\theta}{\diff r}&=&\frac{p_{\bot{\rm eq}}}{r\gamma_{\rm e}}
\\
&\approx&\frac{1}{r\gammaw}
\eqe
for $p_{\bot{\rm eq}}\ge 1$. Integrating over the entire acceleration
phase, one finds an accumulated displacement of 
$\Delta\theta=2\kappaep r_{\rm L}/r_{\rm MHD}$. 
A uniform flow requires $\Omega>\Delta\theta^2$, 
which corresponds to $\Omega\gtrsim \gamma_{{\rm w}0}^{-2}$. Thus, the angular displacement 
or spreading of the beam is small and a uniform, radial flow is a good approximation provided the 
conditions at launch are homogeneous on angular scales smaller than 
${\cal M}^{-1}\sigma_0^{-1}$.

The situation is different in models which extract energy from the
static component of the magnetic field, such as that that proposed by
\citet{bell92}, or the closely related \lq\lq unipolar inductor\rq\rq\
model \citep{blasietal00,arons03}.  There, the electric field
responsible for acceleration can be described by an electrostatic
potential with conical equipotential surfaces of constant latitude. Therefore,
particles are accelerated only if their trajectories migrate in
latitude, and the Hillas limit is reached by charges that drift all
the way from the equator to the pole (or vice-versa). These
models have been developed by computing test-particle trajectories
in the fields given by solutions of the force-free
equations. So far, they do not address the back reaction of the charge
separation and currents implied by the particle flows on the original
fields. Therefore, unlike the solutions presented in Fig.~\ref{figone}, they 
do not provide a self-consistent explanation of how Poynting flux is converted into
the power carried by accelerated particles.

\section{Conclusions}
\label{conclusions}
Inductive acceleration in a Poynting-flux dominated, relativistic
outflow is a viable mechanism for accelerating not only leptons, but
also ions up to energies close to the limiting value given by
\cite{hillas84}. The solutions given in \S~\ref{solutions}, and
illustrated in Fig.~\ref{figone}, explicitly describe the rate of
particle acceleration and the accompanying depletion of Poynting flux,
and demonstrate that, in the presence of ions, comparable power is
channeled into the leptonic and ionic components. 
These solutions provide a
self-consistent theoretical basis for the idea that
objects such as newly born magnetars and pulsars 
are potential sources of ultra-high energy cosmic rays, an hypothesis
whose implications were explored in detail by
\citet{blasietal00} and \citet{arons03}.

The leptonic version of inductive acceleration has previously been
applied to both blazar jets \citep{kirkmochol11} and pulsar wind
nebulae \citep{kirkgiacinti17}. In each case, the primary limitation
on the maximum possible particle energy is imposed by the finite size
of the region traversed by the unperturbed relativistic flow, before
it is terminated by the surrounding medium. Introducing ions into the
flow exerts a confining effect on the leptons, enabling them to reach
a given energy in a much shorter distance. Thus, in the case of
blazars, the presence of ions moves the location of the acceleration
zone in from $\sim 1\,$pc to $\sim 0.1\,$pc, placing it inside the
region where broad emission lines are thought to originate. In the
case of the pulsar wind nebulae that surround the Crab pulsar and the
powerful pulsars B0450$-$69 and J0537-6910, the presence of ions
raises the expected energy with which leptons can be injected into the
nebula by roughly one order of magnitude. In contrast to the related
unipolar inductor model, both purely leptonic flows and flows
containing ions remain tightly collimated in the radial direction
during the acceleration process. 
Radiation losses via synchrotron or jitter radiation are shown 
in \S~\ref{radiationlosses} to be unimportant 
for the dynamics of the flow except, perhaps, in the case of protomagnetars.
Losses by inverse Compton scattering off ambient photons may be more important, depending on
the environment in which the flow is accelerating, 
but we leave a discussion of the observational implications of
such effects to future work.

Inductive acceleration is based on the assumption that the energy
contained in the magnetic fluctuations carried by an expanding outflow
is channeled directly into bulk kinetic energy, rather than first
being released by a dissipative mechanism such as magnetic
reconnection. In \S~\ref{dissipation} it is shown that dissipation is
too slow to be important in flows with an electron multiplicity
$\kappa_{\rm e}\lesssim 10^5 \left(4\pi
  L_{38}/\Omega\right)^{1/4}/\textrm{Max}
\left(\ionfraction^{1/2},1\right)$, in which case inductive
acceleration indeed dominates.  Although this expression is derived in
spherical geometry, the generic nature of the mechanism suggest that a
similar limit applies to other low-density, Poynting-dominated, expanding
flows, such as those thought to be present 
in the jets of active galactic nuclei and in gamma-ray bursts sources.

\acknowledgments

We thank Damien B\'egu\'e, 
Uri Keshet, Yuri Lyubarsky and Brian Reville for helpful discussions. This research was
supported by a Grant from the GIF, the German-Israeli Foundation for
Scientific Research and Development.

\software{Wolfram Research, Inc., Mathematica, Version 11.0, Champaign, IL (2016)}
\bibliographystyle{aasjournal}
\bibliography{references}

\end{document}